\documentclass[conference]{IEEEtran}
\usepackage{graphics, epsfig, amsmath, amsfonts, amssymb, eucal, bm, ulem, textcomp, multirow, graphicx}
\newtheorem{thm}{Theorem}

\newtheorem{lem}{Lemma}
\newtheorem{prop}{Proposition}

\DeclareMathOperator{\tr}{tr}
\newcommand{\reals}{\mathbb{R}}
\newtheorem{example}{Example}

\DeclareMathAlphabet{\eurm}{U}{eur}{m}{n}
\DeclareMathAlphabet{\mathbsf}{OT1}{cmss}{bx}{n}
\DeclareMathAlphabet{\mathssf}{OT1}{cmss}{m}{sl}
\DeclareMathAlphabet{\mathcsf}{OT1}{cmss}{sbc}{n}

\newcommand{\randomvalue}[1]{\eurm{\uppercase{#1}}}

\DeclareSymbolFont{bsfletters}{OT1}{cmss}{bx}{n}  
\DeclareSymbolFont{ssfletters}{OT1}{cmss}{m}{n}
\DeclareMathSymbol{\bsfGamma}{0}{bsfletters}{'000}
\DeclareMathSymbol{\ssfGamma}{0}{ssfletters}{'000}
\DeclareMathSymbol{\bsfDelta}{0}{bsfletters}{'001}
\DeclareMathSymbol{\ssfDelta}{0}{ssfletters}{'001}
\DeclareMathSymbol{\bsfTheta}{0}{bsfletters}{'002}
\DeclareMathSymbol{\ssfTheta}{0}{ssfletters}{'002}
\DeclareMathSymbol{\bsfLambda}{0}{bsfletters}{'003}
\DeclareMathSymbol{\ssfLambda}{0}{ssfletters}{'003}
\DeclareMathSymbol{\bsfXi}{0}{bsfletters}{'004}
\DeclareMathSymbol{\ssfXi}{0}{ssfletters}{'004}
\DeclareMathSymbol{\bsfPi}{0}{bsfletters}{'005}
\DeclareMathSymbol{\ssfPi}{0}{ssfletters}{'005}
\DeclareMathSymbol{\bsfSigma}{0}{bsfletters}{'006}
\DeclareMathSymbol{\ssfSigma}{0}{ssfletters}{'006}
\DeclareMathSymbol{\bsfUpsilon}{0}{bsfletters}{'007}
\DeclareMathSymbol{\ssfUpsilon}{0}{ssfletters}{'007}
\DeclareMathSymbol{\bsfPhi}{0}{bsfletters}{'010}
\DeclareMathSymbol{\ssfPhi}{0}{ssfletters}{'010}
\DeclareMathSymbol{\bsfPsi}{0}{bsfletters}{'011}
\DeclareMathSymbol{\ssfPsi}{0}{ssfletters}{'011}
\DeclareMathSymbol{\bsfOmega}{0}{bsfletters}{'012}
\DeclareMathSymbol{\ssfOmega}{0}{ssfletters}{'012}

\newcommand{\rvs}{{\randomvalue{s}}}	

\newcommand{\rvw}{{\randomvalue{w}}}	

\newcommand{\rvx}{{\randomvalue{x}}}	

\newcommand{\rvz}{{\randomvalue{z}}}	

\begin{document}

\title{A Rate-Distortion Framework for Characterizing Semantic Information}

\author{\authorblockN{Jiakun Liu, Wenyi Zhang}
\authorblockA{Department of Electronic Engineering and Information Science\\
University of Science and Technology of China, Hefei, China\\
liujk@mail.ustc.edu.cn, wenyizha@ustc.edu.cn}
\and
\authorblockN{H. Vincent Poor}
\authorblockA{Department of Electrical Engineering\\
Princeton University, Princeton, NJ, USA\\
poor@princeton.edu}
}

\maketitle

\begin{abstract}
A rate-distortion problem motivated by the consideration of semantic information is formulated and solved. The starting point is to model an information source as a pair consisting of an intrinsic state which is not observable, corresponding to the semantic aspect of the source, and an extrinsic observation which is subject to lossy source coding. The proposed rate-distortion problem seeks a description of the information source, via encoding the extrinsic observation, under two distortion constraints, one for the intrinsic state and the other for the extrinsic observation. The corresponding state-observation rate-distortion function is obtained, and a few case studies of Gaussian intrinsic state estimation and binary intrinsic state classification are studied.
\end{abstract}

\section{Introduction}
\label{sec:intro}

In his landmark paper \cite{shannon48}, Shannon explicitly excluded semantic aspects from his framework of information theory, saying ``these semantic aspects of communication are irrelevant to the engineering problem.'' This exclusion has been followed throughout the development of the core of information theory; see, e.g., \cite{cover06}. Nevertheless, efforts to  characterize semantic aspects of messages and incorporate them into information processing and transmission have been pursued since the inception of information theory; see, e.g., \cite{carnap53} \cite{floridi04} \cite{bao11} \cite{juba11} for a few representative works that cover an extensive range of issues in this context.

In this work, we propose an information theoretic model, motivated by the consideration of semantic information, which has gained much interest recently in the development of 5G and beyond wireless systems \cite{popovski19} \cite{kountouris20}. Instead of seeking a task-independent universal characterization of semantic information, which still appears elusive, we argue that, for many applications the semantic aspects of information correspond to the accomplishment of certain inference goals. So by semantic information, we actually mean that there exists some intrinsic state (i.e., ``feature'') embedded in the sensed extrinsic observation (i.e., ``appearance''), and the interest of the destination is not merely the extrinsic observation, but also the intrinsic state. Hence, if we consider an information theoretic characterization of such a ``semantic'' information source, the task of coding is to efficiently encode the extrinsic observation so that the decoder can infer both the intrinsic state and the extrinsic observation, subject to fidelity criteria on both, simultaneously.

As related topics, the information bottleneck \cite{tishby99} \cite{goldfeld20} and the privacy funnel \cite{makhdoumi14} \cite{shkel21} are, in a certain sense, dual concepts, and both place constraints in terms of information measures. Task-based compression has been tackled mainly from the perspective of quantizer design \cite{shlezinger19}. It has been demonstrated that steering the design goal according to the task leads to performance benefits compared with conventional task-agnostic approach, a conclusion in line with what we advocate in our work. The perception-distortion tradeoff \cite{blau19} imposes an additional constraint on the probability distribution of the reproduction. None of these related works proposes to decompose the information source into intrinsic and extrinsic parts as in our work, let alone investigate the joint behavior of them. In \cite{kipnis21}, a similar intrinsic state-extrinsic observation model is studied, but the encoder is designed based on the marginal distribution of the extrinsic observation only.

We describe the proposed problem formulation in Section \ref{sec:formulation}. We then recognize the proposed problem as a lossy source coding problem with two distortion constraints, one of which is with respect to the unobservable intrinsic state and its reproduction. This problem thus can be cast as an instance of the so-called ``indirect rate-distortion problem'' \cite{dorushin62} \cite{wolf70} \cite{berger71} \cite{witsenhausen80}. We present the corresponding rate-distortion function in Section \ref{sec:tradeoff}. We then investigate several case studies when the intrinsic state and the extrinsic observation are jointly Gaussian, and when the intrinsic state is Bernoulli with conditionally Gaussian extrinsic observations, in Section \ref{sec:case:gaussian} and Section \ref{sec:case:classification}, respectively. Finally, we conclude the paper in Section \ref{sec:conclusion}.

\section{Problem Formulation}
\label{sec:formulation}

\begin{figure}[t]
    \centering
    \includegraphics[width=2.5in]{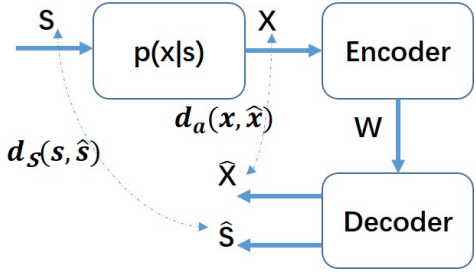}
    \caption{Illustration of system model.}
    \label{fig:semantic-model}
\end{figure}

The mathematical problem formulation is as follows; see Figure \ref{fig:semantic-model} for an illustration. We describe a memoryless information source as a tuple of random variables, $(\rvs, \rvx)$ with joint probability distribution $p(s, x)$ in product alphabet $\mathcal{S} \times \mathcal{X}$. We interpret $\rvs$ as the intrinsic state, which captures the ``semantic'' aspect of the source and is not observable, and $\rvx$ as the extrinsic observation of the source, which captures the ``appearance'' of the source to an observer.

For a length-$n$ independent and identically distributed (i.i.d.) sequence from the source, $(\rvs^n, \rvx^n)$, a source encoder $f_n$ of rate $R$ is a mapping that maps $\rvx^n$ into an index $\rvw$ within $\{1, 2, \ldots, 2^{nR}\}$, and a corresponding decoder $g_n$ is a mapping that maps $\rvw$ into a pair $(\hat{\rvs}^n, \hat{\rvx}^n)$ drawn values from product alphabet $\hat{\mathcal{S}} \times \hat{\mathcal{X}}$. We consider two distortion metrics, $d_\mathrm{s}(s, \hat{s}): \mathcal{S} \times \hat{\mathcal{S}} \mapsto \mathbb{R}^+ \cup \{0\}$ that models the semantic distortion, and $d_\mathrm{a}(x, \hat{x}): \mathcal{X} \times \hat{\mathcal{X}} \mapsto \mathbb{R}^+ \cup \{0\}$ that models the appearance distortion, respectively. So the block-wise distortions are
\begin{eqnarray}
    d_\mathrm{s}(s^n, \hat{s}^n) = \frac{1}{n} \sum_{i = 1}^n d_\mathrm{s}(s_i, \hat{s}_i),\\
    d_\mathrm{a}(x^n, \hat{x}^n) = \frac{1}{n} \sum_{i = 1}^n d_\mathrm{a}(x_i, \hat{x}_i),
\end{eqnarray}
respectively.

For example, the intrinsic state may be categorical, as the label for certain classification task, and the extrinsic observation may be an image or video clip whose content reflects and depends upon the intrinsic state. In applications, a remote viewer may be interested in the extrinsic observation (i.e., image or video clip) itself, whereas another remote pattern classifier may instead be interested in inferring the intrinsic state (i.e., the label) from the encoded extrinsic observation.\footnote{Note that in general the intrinsic state is not a deterministic function of, and hence cannot be perfectly recovered from, the extrinsic observation; see, e.g., \cite[Chap. 2 and 3]{shalev-shwartz}.}

We say that a rate-distortion triple $(R, D_\mathrm{s}, D_\mathrm{a})$ is achievable if there exists a sequence of encoders $\{f_n\}$ and decoders $\{g_n\}$ at rate $R$ such that as $n$ grows without bound, the expected distortions satisfy
\begin{eqnarray}
    \label{eqn:semantic-distortion-constraint}
    \lim_{n \rightarrow \infty} \mathbf{E} d_\mathrm{s}(\rvs^n, \hat{\rvs}^n) &\leq& D_\mathrm{s},\\
    \lim_{n \rightarrow \infty} \mathbf{E} d_\mathrm{a}(\rvx^n, \hat{\rvx}^n) &\leq& D_\mathrm{a}.
\end{eqnarray}
The boundary of the set of all achievable rate-distortion triples is defined as the state-observation rate-distortion function (SORDF).

\section{Characterization of SORDF}
\label{sec:tradeoff}

The following theorem characterizes the SORDF.

\begin{thm}
    \label{thm:rate-distortion function}
    The SORDF of the problem setup considered in Section \ref{sec:formulation} is
    \begin{eqnarray}
        \label{eqn:rate-distortion function}
        R(D_\mathrm{s}, D_\mathrm{a}) &=& \min I(\rvx; \hat{\rvs}, \hat{\rvx}),\\
        \mbox{s.t.}\quad \mathbf{E} \hat{d}_\mathrm{s}(\rvx, \hat{\rvs}) &\leq& D_\mathrm{s},\\
        \mathbf{E} d_\mathrm{a}(\rvx, \hat{\rvx}) &\leq& D_\mathrm{a},\\
        \label{eqn:d-hat}
        \mbox{where}\;\hat{d}_\mathrm{s}(x, \hat{s}) &=& \frac{1}{p(x)} \sum_{s \in \mathcal{S}} p(s, x) d_\mathrm{s}(s, \hat{s}).
    \end{eqnarray}
\end{thm}

\textit{Proof:} The SORDF (\ref{eqn:rate-distortion function}) is basically a combination of the indirect rate-distortion function \cite{dorushin62} \cite{wolf70} \cite[Chap. 3, Sec. 5]{berger71} \cite{witsenhausen80} and the rate-distortion function with multiple distortion constraints \cite[Sec. VII]{elgamal82} \cite[Prob. 7.14]{csiszar} \cite[Prob. 10.19]{cover06}. Hence we only give a sketch of its proof.

A general and unified approach to the indirect rate-distortion function, as adopted in \cite{witsenhausen80}, is first showing that the one-shot expected distortion $\mathbf{E} d_\mathrm{s}(\rvs, \hat{\rvs})$ is equivalent to $\mathbf{E} \hat{d}_\mathrm{s}(\rvx, \hat{\rvs})$, and then invoking a tensorization argument to extend the one-shot equivalence to block codes. Here we combine these two steps, to show that for an arbitrary encoder-decoder pair, the original semantic distortion constraint (\ref{eqn:semantic-distortion-constraint}) is equivalent to
\begin{eqnarray}
    \lim_{n \rightarrow \infty} \mathbf{E} \hat{d}_\mathrm{s}(\rvx^n, \hat{\rvs}^n) &\leq& D_\mathrm{s},
\end{eqnarray}
where $\hat{d}_\mathrm{s}(x^n, \hat{s}^n) = \frac{1}{n} \sum_{i = 1}^n \hat{d}_\mathrm{s} (x_i, \hat{s}_i)$. To prove this, consider an arbitrary encoder-decoder pair, for which it holds that
\begin{eqnarray}
    \mathbf{E} d_\mathrm{s}(\rvs^n, \hat{\rvs}^n) &=& \sum_{s^n, \hat{s}^n} p(s^n, \hat{s}^n) d_\mathrm{s}(s^n, \hat{s}^n)\nonumber\\
    &=& \sum_{s^n, x^n, \hat{s}^n} p(s^n, x^n, \hat{s}^n) d_\mathrm{s}(s^n, \hat{s}^n)\nonumber\\
    &\stackrel{(a)}{=}& \sum_{s^n, x^n, \hat{s}^n} p(s^n | x^n) p(x^n, \hat{s}^n) d_\mathrm{s}(s^n, \hat{s}^n)\nonumber\\
    &=& \sum_{x^n, \hat{s}^n} p(x^n, \hat{s}^n) \sum_{s^n} p(s^n | x^n) d_\mathrm{s}(s^n, \hat{s}^n)\nonumber\\
    &\stackrel{(b)}{=}& \sum_{x^n, \hat{s}^n} p(x^n, \hat{s}^n) \frac{1}{n} \sum_{i = 1}^n \hat{d}_\mathrm{s}(x_i, \hat{s}_i),
\end{eqnarray}
where (a) is due to the Markov chain relationship $\rvs^n \leftrightarrow \rvx^n \leftrightarrow \hat{\rvs}^n$, and (b) is due to the i.i.d. property of $(\rvs^n, \rvx^n)$ and the definition of $\hat{d}_\mathrm{s}$ in (\ref{eqn:d-hat}).

Subsequently, the problem is reduced into a standard lossy source coding problem with multiple distortion constraints \cite[Sec. VII]{elgamal82} \cite[Prob. 7.14]{csiszar} \cite[Prob. 10.19]{cover06}, and the SORDF (\ref{eqn:rate-distortion function}) follows from standard achievability and converse proof techniques. $\Box$

Similar to rate-distortion functions with a single distortion constraint, we have the following properties of $R(D_\mathrm{s}, D_\mathrm{a})$.

\begin{prop}
    \label{prop:properties}
    \begin{enumerate}
        \item $R(D_\mathrm{s}, D_\mathrm{a})$ is monotonically nonincreasing with $D_\mathrm{s}$ and $D_\mathrm{a}$.
        \item $R(D_\mathrm{s}, D_\mathrm{a})$ is jointly convex with respect to $(D_\mathrm{s}, D_\mathrm{a})$.
        \item For any rate $R \geq 0$, the contour set of $(D_\mathrm{s}, D_\mathrm{a})$ such that $R(D_\mathrm{s}, D_\mathrm{a}) \leq R$ is convex.
    \end{enumerate}
\end{prop}

The proof of the first two properties is exactly the same as that for standard rate-distortion functions \cite{berger71, cover06}, and the third property is an immediate corollary of the second property.

\section{Case Study: Jointly Gaussian Model}
\label{sec:case:gaussian}

Consider the case where $\rvs$ and $\rvx$ are jointly Gaussian vectors with zero mean and covariance matrix
\begin{eqnarray}
    \begin{bmatrix}
        \mathbf{\Sigma}_\rvs & \mathbf{\Sigma}_{\rvs \rvx} \\
        \mathbf{\Sigma}_{\rvs \rvx}^T & \mathbf{\Sigma}_\rvx
    \end{bmatrix},
\end{eqnarray}
and the distortion metrics are squared error, as $d_\mathrm{s}(s, \hat{s}) = \|s - \hat{s}\|^2$ and $d_\mathrm{a}(x, \hat{x}) = \|x - \hat{x}\|^2$ respectively.

Note that conditioned upon $x$, $\rvs$ is conditionally Gaussian as $\rvs|x \sim \mathcal{N}\left(\mathbf{\Sigma}_{\rvs \rvx} \mathbf{\Sigma}_\rvx^{-1} x, \mathbf{\Sigma}_\rvs - \mathbf{\Sigma}_{\rvs \rvx} \mathbf{\Sigma}_\rvx^{-1} \mathbf{\Sigma}_{\rvs \rvx}^T \right)$. So the equivalent semantic distortion metric is
\begin{eqnarray}
    &&\hat{d}_\mathrm{s}(x, \hat{s}) = \mathbf{E}_{\rvs|x} \|\rvs - \hat{s}\|^2\nonumber\\
    &=& \mathrm{tr} \left[\mathbf{\Sigma}_\rvs - \mathbf{\Sigma}_{\rvs \rvx} \mathbf{\Sigma}_\rvx^{-1} \mathbf{\Sigma}_{\rvs \rvx}^T\right] + \|\mathbf{\Sigma}_{\rvs \rvx} \mathbf{\Sigma}_\rvx^{-1} x - \hat{s}\|^2,
\end{eqnarray}
where the first trace term is exactly the minimum mean-squared error (MMSE) of estimating $\rvs$ upon observing $\rvx$, denoted as $\mathsf{mmse}$ in the sequel. The SORDF hence becomes
\begin{eqnarray}
    \label{eqn:SORDF-Gaussian-general}
    R(D_\mathrm{s}, D_\mathrm{a}) &=& \min I(\rvx; \hat{\rvs}, \hat{\rvx}),\\
    \label{eqn:SORDF-Gaussian-general-Ds}
    \mbox{s.t.}\quad \mathbf{E} \|\mathbf{\Sigma}_{\rvs \rvx} \mathbf{\Sigma}_\rvx^{-1} \rvx - \hat{\rvs}\|^2 &\leq& D_\mathrm{s} - \mathsf{mmse},\\
    \label{eqn:SORDF-Gaussian-general-Da}
    \mathbf{E} \|\rvx - \hat{\rvx}\|^2 &\leq& D_\mathrm{a}.
\end{eqnarray}

\subsection{$R(\infty, D_\mathrm{a})$ and $R(D_\mathrm{s}, \infty)$}
\label{subsec:Gaussian-degenerated}

Without the semantic distortion constraint, i.e., $D_\mathrm{s} = \infty$, $R(\infty, D_\mathrm{a})$ is the well known rate-distortion function for vector Gaussian source; alternatively, without the appearance distortion constraint, i.e., $D_\mathrm{a} = \infty$, $R(D_\mathrm{s}, \infty)$ is given by the following result, which generalizes the case studied in \cite{wolf70} to jointly Gaussian vectors.

\begin{prop}
    \label{prop:gaussian-semantic-only}
    For the jointly Gaussian source model, $R(D_\mathrm{s}, \infty) = R_1(D_\mathrm{s} - \mathsf{mmse})$, where $R_1(D)$ is the rate-distortion function for $\mathbf{\Sigma}_{\rvs \rvx} \mathbf{\Sigma}_\rvx^{-1} \rvx$ under the squared error distortion metric.
\end{prop}

\textit{Proof:} We consider an estimate-and-compress scheme which first transforms the source observation $\rvx$ into $\mathbf{\Sigma}_{\rvs \rvx} \mathbf{\Sigma}_\rvx^{-1} \rvx$, and then encodes the transformed source observation under mean squared error distortion constraint $D_\mathrm{s} - \mathsf{mmse}$. The resulting achievable rate $R_1(D_\mathrm{s} - \mathsf{mmse})$ hence constitutes an upper bound of $R(D_\mathrm{s}, \infty)$.

To show that the scheme described above is indeed optimal, consider any $\hat{\rvs}$ jointly distributed with $\rvx$, satisfying the distortion constraint. Note that $\mathbf{\Sigma}_{\rvs \rvx} \mathbf{\Sigma}_\rvx^{-1} \rvx \leftrightarrow \rvx \leftrightarrow \hat{\rvs}$ constitute a Markov chain. So according to the data processing inequality, $I(\mathbf{\Sigma}_{\rvs \rvx} \mathbf{\Sigma}_\rvx^{-1} \rvx; \hat{\rvs}) \leq I(\rvx; \hat{\rvs})$. Since this holds for any $\hat{\rvs}$, it holds when $I(\rvx; \hat{\rvs}) = R(D_\mathrm{s}, \infty)$, and thus
\begin{eqnarray*}
    R_1(D_\mathrm{s} - \mathsf{mmse}) \leq I(\mathbf{\Sigma}_{\rvs \rvx} \mathbf{\Sigma}_\rvx^{-1} \rvx; \hat{\rvs}) \leq I(\rvx; \hat{\rvs}) = R(D_\mathrm{s}, \infty),
\end{eqnarray*}
thereby completing the proof. $\Box$

\subsection{Scalar Case}
\label{subsec:Gaussian-scalar}

Then we consider the evaluation of $R(D_\mathrm{s}, D_\mathrm{a})$ for the special case where both $\rvs$ and $\rvx$ are scalar. Hence $\mathbf{\Sigma}_\rvs, \mathbf{\Sigma}_\rvx$ and $\mathbf{\Sigma}_{\rvs \rvx}$ are all scalar-valued, and $\mathsf{mmse} = \mathbf{\Sigma}_\rvs - \mathbf{\Sigma}_{\rvs \rvx}^2/\mathbf{\Sigma}_\rvx$. We have the following result regarding its SORDF.

\begin{prop}
    \label{prop:SORDF-scalar-Gaussian}
    For the jointly Gaussian source model where both $\rvs$ and $\rvx$ are scalar, its SORDF is given by
    \begin{eqnarray}
        \label{eqn:SORDF-scalar-Gaussian}
        R(D_\mathrm{s}, D_\mathrm{a}) &=& \frac{1}{2} \max\left\{\left(\log \frac{\mathbf{\Sigma}_\rvx}{D_\mathrm{a}}\right)^+,\right.\nonumber\\
        &&\left. \left(\log \frac{\mathbf{\Sigma}_{\rvs \rvx}^2}{\mathbf{\Sigma}_\rvx (D_\mathrm{s} - \mathsf{mmse})}\right)^+ \right\},
    \end{eqnarray}
    for $D_\mathrm{s} > \mathsf{mmse}$, $D_\mathrm{a} > 0$, where $(x)^+$ denotes $\max\{x, 0\}$.
\end{prop}

\textit{Proof:} To show the converse, we note that $R(D_\mathrm{s}, D_\mathrm{a})$ (\ref{eqn:SORDF-Gaussian-general}) is lower bounded by both $\min I(\rvx; \hat{\rvs})$ under (\ref{eqn:SORDF-Gaussian-general-Ds}) and $\min I(\rvx; \hat{\rvx})$ under (\ref{eqn:SORDF-Gaussian-general-Da}). So the first term in the max operand of (\ref{eqn:SORDF-scalar-Gaussian}) is due to the standard Gaussian rate-distortion function, and the second term is due to Proposition \ref{prop:gaussian-semantic-only}.

To show the achievability, we consider two situations. First, if $D_\mathrm{a}/\mathbf{\Sigma}_\rvx^2 \geq (D_\mathrm{s} - \mathsf{mmse})/\mathbf{\Sigma}_{\rvs \rvx}^2$, we let $(\rvx, \hat{\rvs})$ be generated so as to solve the standard Gaussian rate-distortion problem subject to constraint (\ref{eqn:SORDF-Gaussian-general-Ds}) and hence achieve
\begin{eqnarray}
    I(\rvx; \hat{\rvs}) = \frac{1}{2} \left(\log \frac{\mathbf{\Sigma}_{\rvs \rvx}^2}{\mathbf{\Sigma}_\rvx (D_\mathrm{s} - \mathsf{mmse})} \right)^+.
\end{eqnarray}
We further let $\hat{\rvx} = (\mathbf{\Sigma}_\rvx / \mathbf{\Sigma}_{\rvs \rvx}) \hat{\rvs}$, which then satisfies the constraint (\ref{eqn:SORDF-Gaussian-general-Da}), and leads to $I(\rvx; \hat{\rvs}, \hat{\rvx}) = I(\rvx; \hat{\rvs})$ because $\hat{\rvx} \leftrightarrow \hat{\rvs} \leftrightarrow \rvx$. Alternatively, if $D_\mathrm{a}/\mathbf{\Sigma}_\rvx^2 < (D_\mathrm{s} - \mathsf{mmse})/\mathbf{\Sigma}_{\rvs \rvx}^2$, we let $(\rvx, \hat{\rvx})$ be generated so as to solve the standard Gaussian rate-distortion problem subject to constraint (\ref{eqn:SORDF-Gaussian-general-Da}), and let $\hat{\rvs} = (\mathbf{\Sigma}_{\rvs \rvx}/\mathbf{\Sigma}_\rvx) \hat{\rvx}$. These then satisfy constraints (\ref{eqn:SORDF-Gaussian-general-Ds}) and (\ref{eqn:SORDF-Gaussian-general-Da}), and achieve
\begin{eqnarray}
    I(\rvx; \hat{\rvs}, \hat{\rvx}) = I(\rvx; \hat{\rvx}) = \frac{1}{2} \left(\log \frac{\mathbf{\Sigma}_\rvx}{D_\mathrm{a}}\right)^+.
\end{eqnarray}
Putting these two situations together establishes the achievability. $\Box$

The interpretation of Proposition \ref{prop:SORDF-scalar-Gaussian} is rather straightforward. Since $\rvs$ and $\rvx$ are both scalar, their ``directions'' are both degenerated and the goals of reproducing them can be viewed as perfectly ``aligned''. In the achievability proof, the first situation arises when $D_\mathrm{s}$ is small, i.e., when reproducing $\rvs$ is more demanding than reproducing $\rvx$, and the second situation arises when the opposite is true. In both situations, however, note that $\hat{\rvx}$ and $\hat{\rvs}$ are proportional with the same proportion.

\subsection{Vector Case}
\label{subsec:Gaussian-vector}

In this subsection we evaluate the SORDF for the special vector case where $\rvs$ is scalar and $\mathbf{\Sigma}_{\rvs \rvx} \mathbf{\Sigma}_\rvx^{-1}$ coincides with one of the eigenvectors of $\mathbf{\Sigma}_\rvx$, and leave the general vector case to Appendix. Consider the following model:
\begin{eqnarray}
    \label{eqn:aligned-Gaussian-case}
    \rvx = \mathbf{1}_m \rvs + \rvz,
\end{eqnarray}
where $\rvs \sim \mathcal{N}(0, \sigma_\rvs^2)$, $\mathbf{1}_m$ is a lengh-$m$ all-one vector, and $\rvz \sim \mathcal{N}(\mathbf{0}, \sigma_\rvz^2 \mathbf{I})$. Denote the MMSE by $\mathsf{mmse} = \frac{\sigma_\rvs^2 \sigma_\rvz^2}{m \sigma_\rvs^2 + \sigma_\rvz^2}$ and set $\alpha = \frac{m \sigma_\rvs^2 + \sigma_\rvz^2}{\sqrt{m} \sigma_\rvs^2}$. We have the following result.

\begin{prop}
    \label{prop:aligned-Gaussian-case}
    For the jointly Gaussian source model (\ref{eqn:aligned-Gaussian-case}), its SORDF is given by:
    
    - if $D_{\mathrm{s}} \geq \mathsf{mmse}$ and $m \alpha^{2} ( D_{\mathrm{s}} - \mathsf{mmse} ) \leq D_{\mathrm{a}} \leq \alpha^{2} ( D_{\mathrm{s}} - \mathsf{mmse} ) + ( m - 1 ) \sigma_\rvz^2$,
        \begin{eqnarray}
        &&R ( D_{\mathrm{s}} , D_{\mathrm{a}} )
        = \frac{1}{2} \log \left(
          \frac{m \sigma_\rvs^2 + \sigma_\rvz^2}
          {\alpha^{2} ( D_{\mathrm{s}} - \mathsf{mmse} )}
        \right)\nonumber\\
        && \quad + \frac{m - 1}{2} \log \left(
          \frac{( m - 1 ) \sigma_\rvz^2}
          {D_{\mathrm{a}} - \alpha^{2} ( D_{\mathrm{s}} - \mathsf{mmse} )}
        \right) ;
        \end{eqnarray}
    
    - if $0 \le D_{\mathrm{a}} < m \sigma_\rvz^2$ and $\alpha^{2} ( D_{\mathrm{s}} - \mathsf{mmse} ) \ge D_{\mathrm{a}} / m$,
        \begin{eqnarray}
        R ( D_{\mathrm{s}} , D_{\mathrm{a}} ) = \frac{1}{2} \log \left(
          \frac{m^{2} \sigma_\rvs^2 + m \sigma_\rvz^2}
          {D_{\mathrm{a}}}
        \right) + \frac{m - 1}{2}
        \log \left( \frac{m \sigma_\rvz^2}{D_{\mathrm{a}}} \right) ;
        \end{eqnarray}

    - if $0 \le \alpha^{2} ( D_{\mathrm{s}} - \mathsf{mmse} ) < m \sigma_\rvs^2 + \sigma_\rvz^2$ and $D_{\mathrm{a}} > \alpha^{2} ( D_{\mathrm{s}} - \mathsf{mmse} ) + ( m - 1 ) \sigma_\rvz^2$,
        \begin{eqnarray}
        R ( D_{\mathrm{s}} , D_{\mathrm{a}} )
        = \frac{1}{2} \log \left(
          \frac{m \sigma_\rvs^2 + \sigma_\rvz^2}
          {\alpha^{2} ( D_{\mathrm{s}} - \mathsf{mmse} )}
        \right) ;
        \end{eqnarray}
    
    - if $m \sigma_\rvz^2 \le D_{\mathrm{a}} < m \sigma_\rvs^2 + m \sigma_\rvz^2$ and $\alpha^{2} ( D_{\mathrm{s}} - \mathsf{mmse} ) \ge D_{\mathrm{a}} - ( m - 1 ) \sigma_\rvz^2$,
        \begin{eqnarray}
        R ( D_{\mathrm{s}} , D_{\mathrm{a}} )
        = \frac{1}{2} \log \left(
          \frac{m \sigma_\rvs^2 + \sigma_\rvz^2}
          {D_{\mathrm{a}} - ( m - 1 ) \sigma_\rvz^2}
        \right) ;
        \end{eqnarray}

    - if $D_{\mathrm{a}} \ge m \sigma_\rvs^2 + m \sigma_\rvz^2$ and $\alpha^{2} ( D_{\mathrm{s}} - \mathsf{mmse} ) \ge m \sigma_\rvs^2 + \sigma_\rvz^2$,
        \begin{eqnarray}
        R ( D_{\mathrm{s}} , D_{\mathrm{a}} )
        = 0 .
        \end{eqnarray}
\end{prop}

\textit{Proof:} We give an outline of the proof. The key observation is that $\mathbf{\Sigma}_{\rvs \rvx} \mathbf{\Sigma}_\rvx^{-1} \propto \mathbf{b}_1 = \frac{1}{\sqrt{m}} \mathbf{1}_m$ is a unit-norm eigenvector of $\mathbf{\Sigma}_\rvx$, associated with the eigenvalue $m\sigma_\rvs^2 + \sigma_\rvz^2$. The remaining $m - 1$ unit-norm eigenvectors of $\mathbf{\Sigma}_\rvx$ are
\begin{eqnarray}
    \mathbf{b}_i = \left[\underbrace{\frac{1}{\sqrt{i(i - 1)}}, \ldots, \frac{1}{\sqrt{i(i - 1)}}}_{i - 1}, -\sqrt{\frac{i - 1}{i}}, \underbrace{0, \ldots, 0}_{m - i}\right]^T,
\end{eqnarray}
for $i = 2, \ldots, m$, all associated with the identical eigenvalue $\sigma_\rvz^2$. So $\mathbf{B} = \left[\mathbf{b}_1, \ldots, \mathbf{b}_m\right]^T$ is an orthonormal matrix that decorrelates $\rvx$. The SORDF problem (\ref{eqn:SORDF-Gaussian-general})-(\ref{eqn:SORDF-Gaussian-general-Da}) can then be equivalently rewritten as
\begin{eqnarray}
R(D_\mathrm{s}, D_\mathrm{a}) &=& \min I(\mathbf{B}\rvx; \alpha \hat{\rvs}, \mathbf{B} \hat{\rvx}),\\
\mbox{s.t.}\quad \mathbf{E} (\mathbf{b}_1^T \rvx - \alpha \hat{\rvs})^2 &\leq& \alpha^2 (D_\mathrm{s} - \mathsf{mmse})\\
\sum_{i = 1}^m \mathbf{E}(\mathbf{b}_i^T \rvx - \mathbf{b}_i^T \hat{\rvx})^2 &\leq& D_\mathrm{a}.    
\end{eqnarray}

Note that the $m$ elements of $\mathbf{B} \rvx$ are now decorrelated to be independent, and hence it can be shown that the minimization of $I(\mathbf{B}\rvx; \alpha \hat{\rvs}, \mathbf{B} \hat{\rvx})$ can be decoupled and converted into a distortion allocation problem similar to that for the standard parallel Gaussian reverse waterfilling \cite{cover06}. The resulting optimization problem becomes
\begin{eqnarray}
    R(D_\mathrm{s}, D_\mathrm{a}) &=& \min_{(D_1, D_2, \ldots, D_m) \in \mathcal{A}(D_\mathrm{s}, D_\mathrm{a})} \left[R\left(\frac{D_1}{m \sigma_\rvs^2 + \sigma_\rvz^2}\right)\right. \nonumber\\
    &&\left.+ \sum_{i = 2}^m R\left(\frac{D_i}{\sigma_\rvz^2}\right)\right],\\
    \mathcal{A}(D_\mathrm{s}, D_\mathrm{a}) &=& \left\{(D_1, D_2, \ldots, D_m): D_1 \leq \alpha^2 (D_\mathrm{s} - \mathsf{mmse}), \right.\nonumber\\
    && \left.\sum_{i = 1}^m D_i \leq D_\mathrm{a}, D_i \geq 0, \forall i \right\},
\end{eqnarray}
where $R(x) = \frac{1}{2} \left(\log \frac{1}{x}\right)^+$ for $x > 0$. Solving this optimization problem, we obtain the SORDF as presented in Proposition \ref{prop:aligned-Gaussian-case}. $\Box$

\begin{figure}[t]
    \centering
    \includegraphics[width=2.7in]{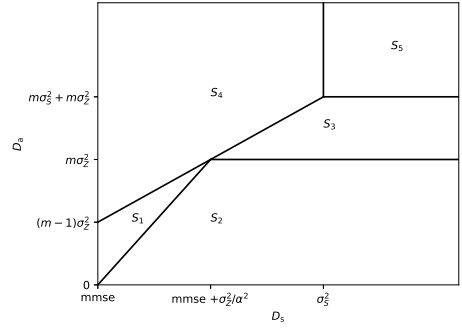}
    \caption{Illustration of the five regions of $(D_\mathrm{s}, D_\mathrm{a})$-plane.}
    \label{fig:regions}
\end{figure}

\begin{figure}[th]
    \centering
    \begin{minipage}{4.2cm}
    \includegraphics[width=4.6cm]{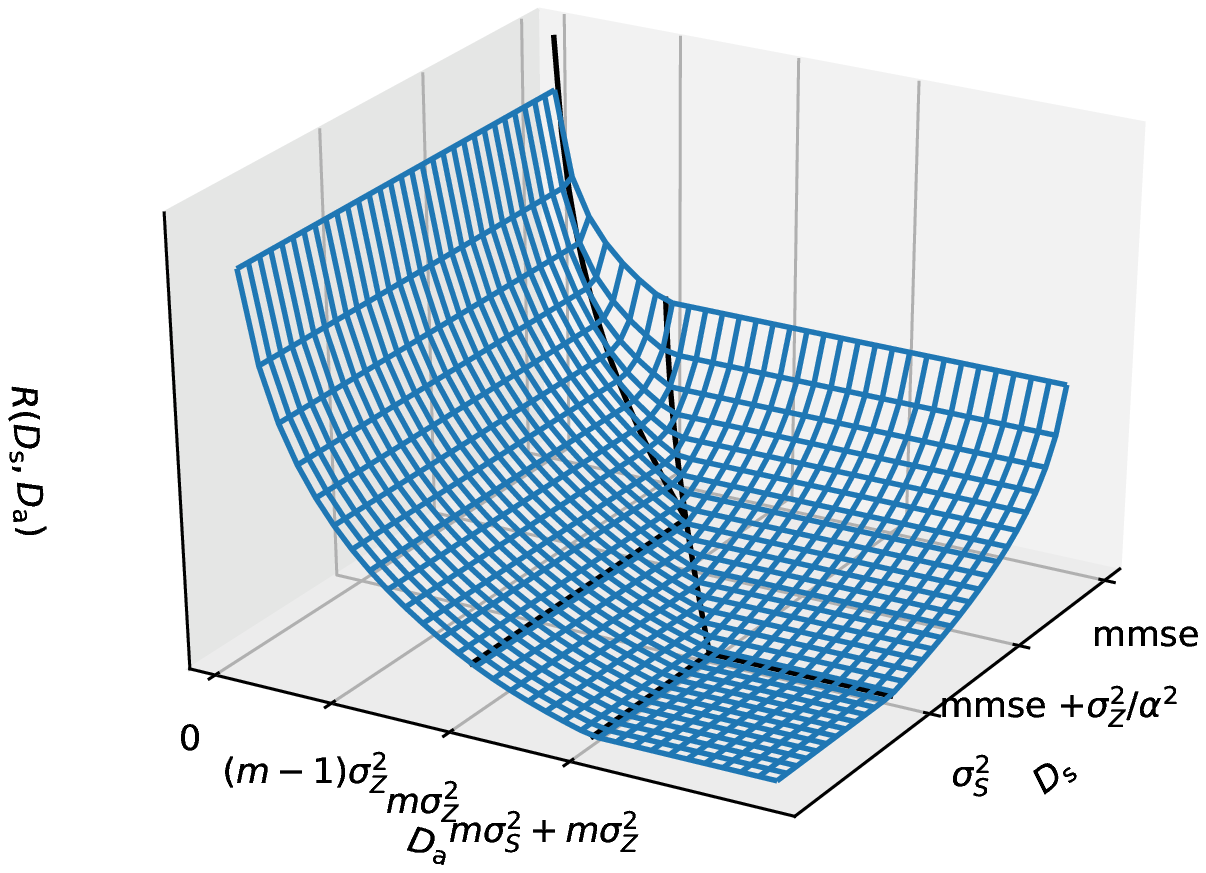}
    \end{minipage}
    \begin{minipage}{4.2cm}
    \includegraphics[width=4.6cm]{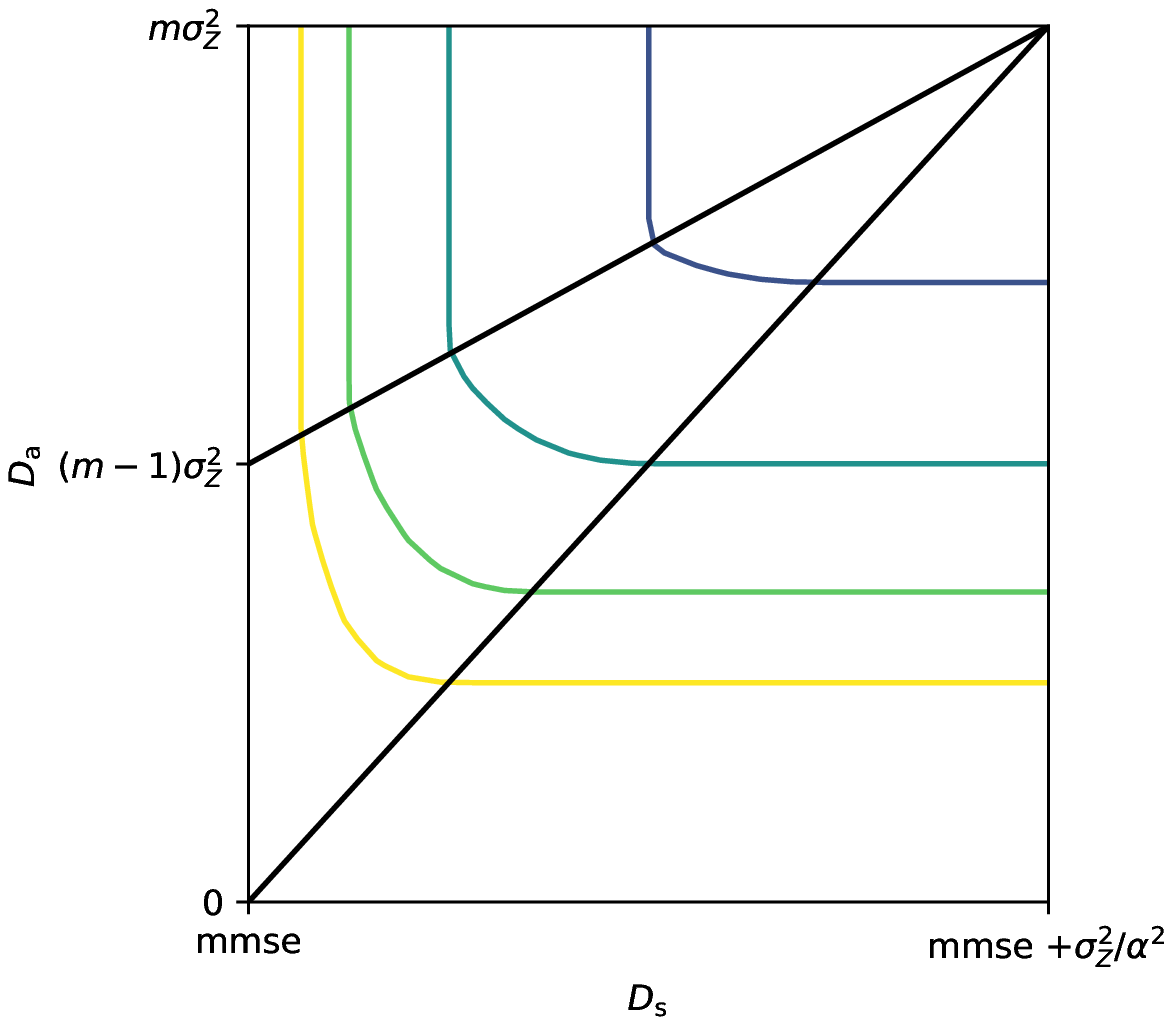}
    \end{minipage}
    \caption{The SORDF $R(D_\mathrm{s}, D_\mathrm{a})$ (left) and its contour plot (right).}
    \label{fig:SORDF-vector-Gaussian}
\end{figure}

According to Proposition \ref{prop:aligned-Gaussian-case}, the $(D_\mathrm{s}, D_\mathrm{a})$-plane is divided into five regions. This is illustrated in Figure \ref{fig:regions}. The SORDF $R(D_\mathrm{s}, D_\mathrm{a})$ and its contour plot are illustrated in Figure \ref{fig:SORDF-vector-Gaussian}. The region where $D_\mathrm{s}$ and $D_\mathrm{a}$ exhibit a tradeoff is clearly indicated by the two slanted-line boundaries in the contour plot: in that region, if we encode $\rvx$ regardless of considering $\rvs$, then extra distortion on $\rvs$ will be incurred, and vice versa.

\section{Case Study: Classification}
\label{sec:case:classification}

Consider the case where $\rvs$ is a binary state, i.e., a Bernoulli random variable drawn from $\{0, 1\}$ with prior probability $1/2$ uniformly. The extrinsic observation $\rvx$ is conditionally Gaussian, as
\begin{eqnarray}
    \label{eqn:source-classification}
    \rvx \sim \mathcal{N}(A, \sigma^2), \quad \mbox{if}\; \rvs = 0;
    \rvx \sim \mathcal{N}(-A, \sigma^2), \quad \mbox{if}\; \rvs = 1. 
\end{eqnarray}
So the marginal distribution of $\rvx$ is a Gaussian mixture. We adopt a Hamming distortion between $\rvs$ and $\hat{\rvs}$, i.e., $d_\mathrm{s}(s, \hat{s}) = 0$ if $s = \hat{s}$ and $1$ otherwise; and a squared error distortion between $\rvx$ and $\hat{\rvx}$, i.e., $d_\mathrm{a}(x, \hat{x}) = (x - \hat{x})^2$.

For this source model, we can obtain its $R(D_\mathrm{s}, \infty)$ in the following result.

\begin{prop}
    \label{prop:SORDF-classification}
    For the source model (\ref{eqn:source-classification}), we have
    \begin{eqnarray}
        R(D_\mathrm{s}, \infty) = 1 - \frac{1}{2}\int_{-\infty}^\infty \left[N^+(x) + N^-(x)\right] h_2(g(x)) \mathrm{d}x,
    \end{eqnarray}
    for $Q(A/\sigma) \leq D_\mathrm{s} \leq 1/2$, and $R(D_\mathrm{s}, \infty) = 0$ for $D_\mathrm{s} > 1/2$, where
    \begin{eqnarray}\label{eqn:optimal-g}
        g(x) = \left[1 + \exp\left(\lambda \frac{1 - e^{-2Ax/\sigma^2}}{1 + e^{-2Ax/\sigma^2}}\right)\right]^{-1},
    \end{eqnarray}
    wherein $\lambda < 0$ is chosen so as to satisfy
    \begin{eqnarray}\label{eqn:optimal-lambda}
        \int_{-\infty}^\infty \left[N^+(x) - N^-(x)\right] g(x) \mathrm{d}x = 1 - 2 D_\mathrm{s}.
    \end{eqnarray}
    Here we denote by $N^+(x)$ and $N^-(x)$ the probability density functions of $\mathcal{N}(A, \sigma^2)$ and $\mathcal{N}(-A, \sigma^2)$, respectively, and $h_2(t)$ is the binary entropy function, $h_2(t) = - t\log_2 t - (1 - t) \log_2 (1 - t)$, for $0 \leq t \leq 1$.
\end{prop}

\textit{Proof:} The expression of $R(D_\mathrm{s}, \infty)$ is obtained by solving $\min I(\rvx; \hat{S})$, subject to the constraint of
\begin{eqnarray}
    \mathbf{E} \hat{d}_\mathrm{s} (\rvx, \hat{\rvs}) \leq D_\mathrm{s},
\end{eqnarray}
by optimizing the conditional probability $g(x) = \mathrm{Pr}(\hat{s} = 0 | x)$, where the expectation $\mathbf{E} \hat{d}_\mathrm{s} (\rvx, \hat{\rvs})$ can be further evaluated as
\begin{eqnarray}
    \frac{1}{2} \int_{-\infty}^\infty \left[N^-(x) g(x) + N^+(x) (1 - g(x))\right] \mathrm{d}x.
\end{eqnarray}

Note that due to the symmetry in the model, the optimal $g(x)$ should satisfy $g(x) + g(-x) = 1$, $\forall x$, and consequently the resulting $\hat{\rvs}$ is uniform Bernoulli. This property is satisfied by (\ref{eqn:optimal-g}). $\Box$

\begin{figure}[th]
    \centering
    \begin{minipage}{4.2cm}
    \includegraphics[width=4.6cm]{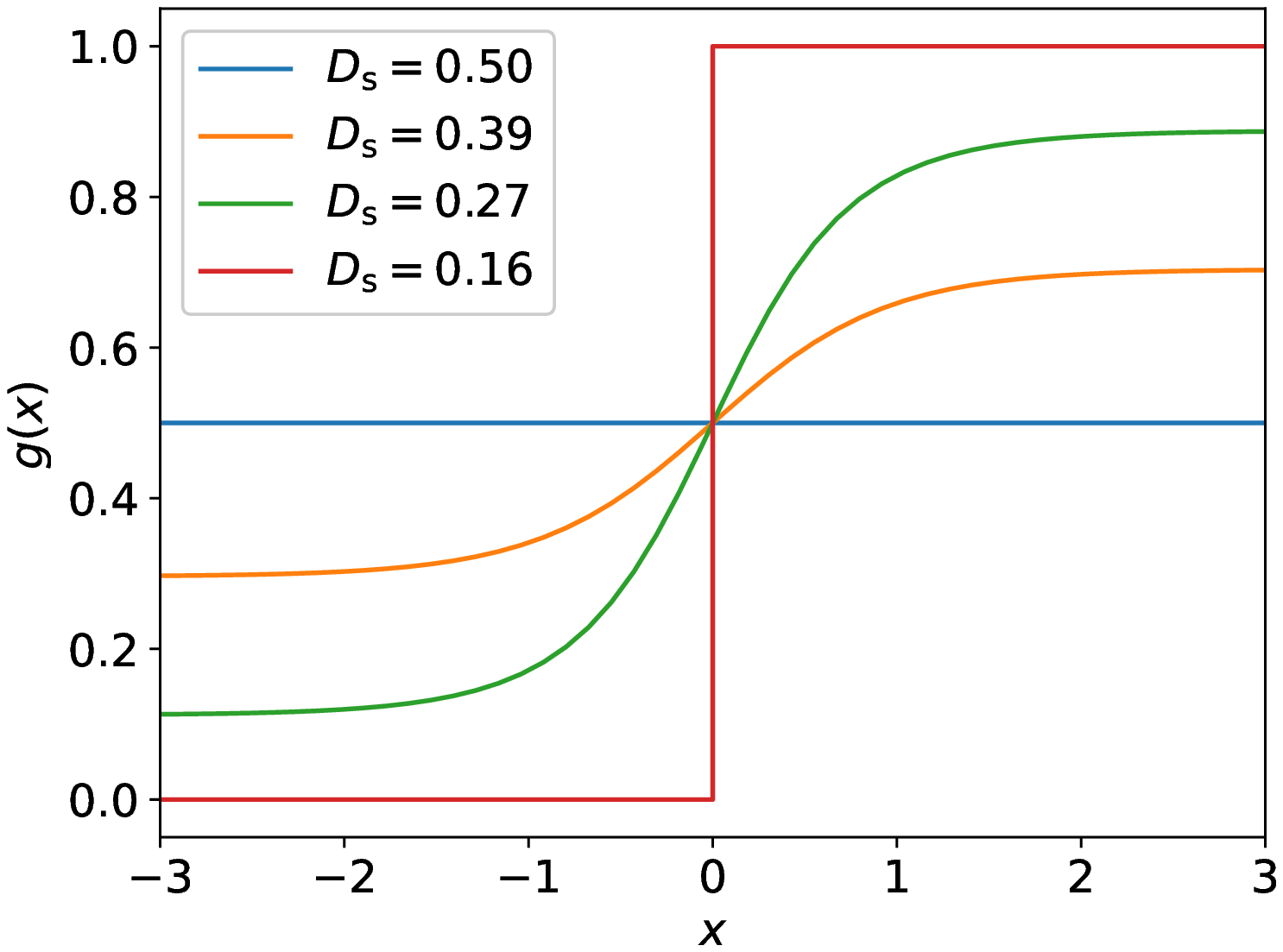}
    \end{minipage}
    \begin{minipage}{4.2cm}
    \includegraphics[width=4.6cm]{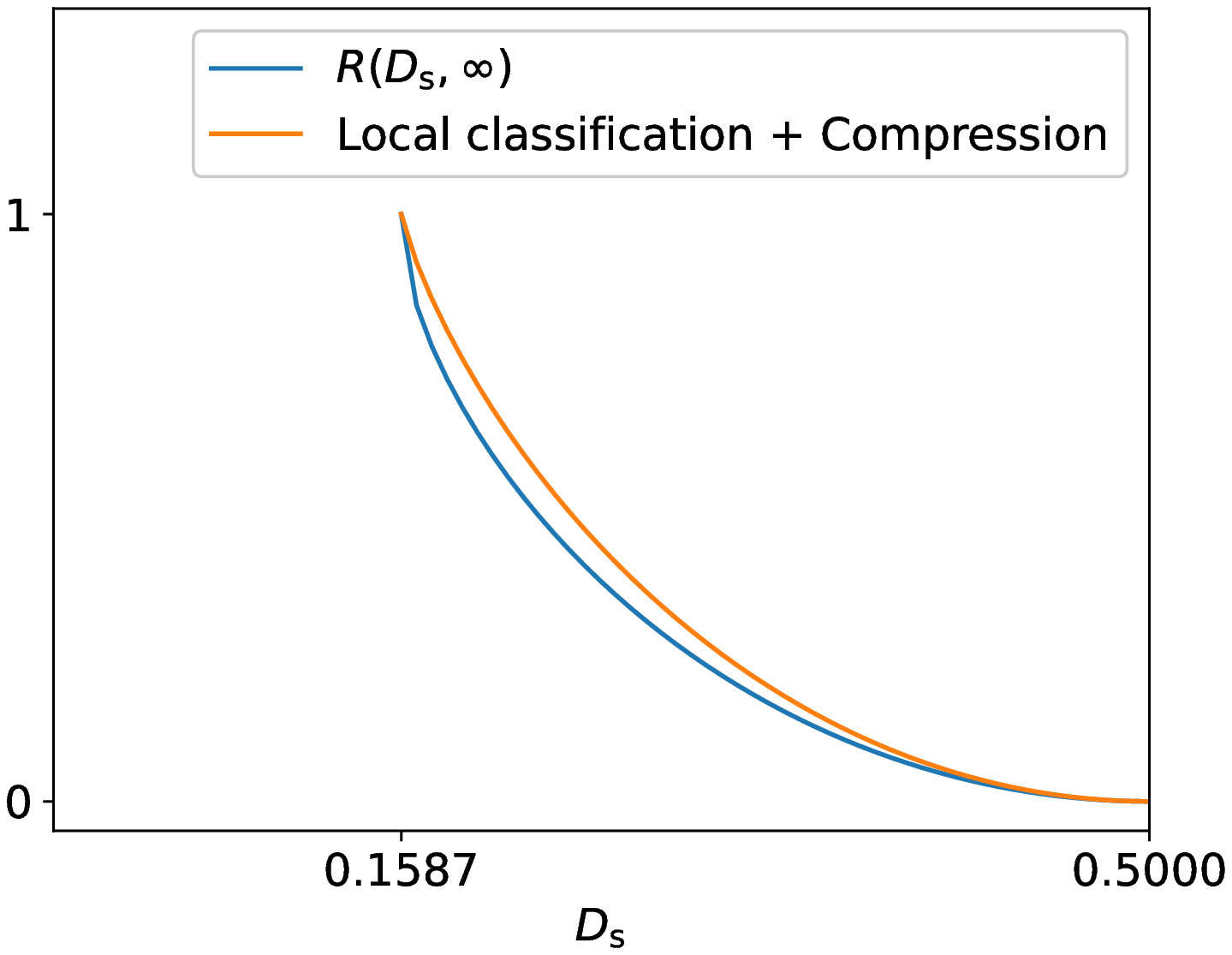}
    \end{minipage}
    \caption{$g(x)$ (left) and the corresponding $R(D_\mathrm{s}, \infty)$ (right), $A = \sigma^2 = 1$, $Q(A/\sigma) = 0.1587$.}
    \label{fig:SORDF-classification-only}
\end{figure}

The conditional probability $g(x)$ as given by (\ref{eqn:optimal-g}) can be interpreted as a soft weighting of the posterior belief regarding $\rvs$ upon observing $\rvx$; see Figure \ref{fig:SORDF-classification-only}. Statistically, observing a positive $x \gg 0$ strongly suggests a possibility of $\rvs = 0$, and thus $g(x)$ is large, while observing a negative $x \ll 0$ leads to the opposite; alternatively, the least informative case of $x \approx 0$ results in $g(x) \approx 1/2$. A noteworthy consequence revealed by Proposition \ref{prop:SORDF-classification} is that the naive scheme of performing locally optimal (Bayesian) classification and encoding the binary classification is suboptimal (indicated as ``Local classification + Compression'' in Figure \ref{fig:SORDF-classification-only}), except for the extreme case of $D_\mathrm{s} = Q(A/\sigma)$. This is different from the jointly Gaussian case in Section \ref{sec:case:gaussian}, where Proposition \ref{prop:gaussian-semantic-only} (see also \cite{wolf70}) indicates the the estimate-and-compress scheme is optimal.

Based upon Proposition \ref{prop:SORDF-classification}, we have the following achievability result.
\begin{prop}
    \label{prop:classification-upper-bound}
    For the source model (\ref{eqn:source-classification}), we have
    \begin{eqnarray}\label{eqn:classification-sordf}
        R(D_\mathrm{s}, D_\mathrm{a})\!\!\!\! &\leq& \!\!\!\!\!\!\!\!\!\!\min_{D \in \left[Q\left(\frac{A}{\sigma}\right), D_\mathrm{s}\right]} \!\!\! \left\{R(D, \infty) + \frac{1}{2} \left(\log \frac{\eta}{D_\mathrm{a}}\right)^+\right\},\\
        \eta &=& \int_{-\infty}^\infty (x - \gamma)^2 \left[N^+(x) + N^-(x)\right] g_D(x) \mathrm{d}x,\nonumber\\
        \gamma &=& \int_{-\infty}^\infty x \left[N^+(x) + N^-(x)\right] g_D(x) \mathrm{d}x\nonumber,
    \end{eqnarray}
    where $g_D(x)$ is given by (\ref{eqn:optimal-g}) satisfying (\ref{eqn:optimal-lambda}) whose right hand side is now replaced by $1 - 2D$.
\end{prop}

\textit{Proof:} Here we give an outline of a coding scheme that leads to the proof of Proposition \ref{prop:classification-upper-bound}. We first apply Proposition \ref{prop:SORDF-classification} to encode $\rvx$ into $\hat{\rvs}$ at rate $R(D, \infty)$ so as to satisfy the semantic distortion constraint $D_\mathrm{s}$, noting that $D \in [Q(A/\sigma), D_\mathrm{s}]$. Then, conditioned upon $\hat{\rvs}$, we encode $\rvx - \mathbf{E} [\rvx | \hat{\rvs}]$ into $\tilde{\rvx}$ using an i.i.d. Gaussian codebook ensemble with mean squared error distortion constraint $D_\mathrm{a}$, which can be successfully accomplished at rate $\frac{1}{2} \left(\log \frac{\eta}{D_\mathrm{a}}\right)^+$ \cite[Thm. 3]{lapidoth97}. Finally, the decoder reproduces $\hat{\rvx} = \tilde{\rvx} + \mathbf{E} [\rvx | \hat{\rvs}]$. Since the aforementioned scheme applies to any $D \in [Q(A/\sigma), D_\mathrm{s}]$, optimizing $D$ leads to (\ref{eqn:classification-sordf}). $\Box$

Figure \ref{prop:classification-upper-bound} displays the achievable upper bound of $R(D_\mathrm{s}, D_\mathrm{a})$ in (\ref{eqn:classification-sordf}). For comparison, we also plot $(1/2) \left[\log \left(\sigma^2/D_\mathrm{a}\right)\right]^+$, which corresponds to the rate-distortion function under the ideal scenario where both the encoder and the decoder know $\rvs$ perfectly, and $(1/2) \left[\log \left((A^2 + \sigma^2)/D_\mathrm{a}\right)\right]^+$, which corresponds to the naive scheme which directly encodes $\rvx$ subject to the squared error distortion with an i.i.d. Gaussian codebook ensemble.

\begin{figure}[th]
    \centering
    \includegraphics[width=2.7in]{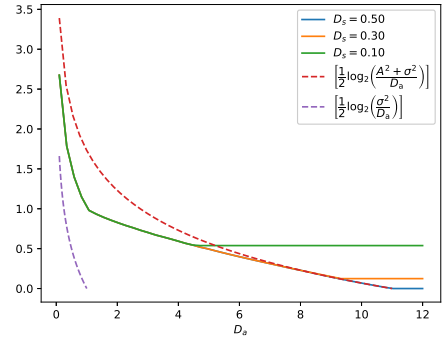}
    \caption{Achievable upper bound of $R(D_\mathrm{s}, D_\mathrm{a})$ in (\ref{eqn:classification-sordf}), $A^2/\sigma^2 = 10$.}
    \label{fig:SORDF-classification-bound}
\end{figure}

\section{Conclusion}
\label{sec:conclusion}

We have provided a general rate-distortion framework for characterizing an information source that can be modeled as a tuple of an intrinsic state and an extrinsic observation. Two issues are particularly relevant for the application of this framework --- first, developing efficient numerical algorithms for computing the SORDF for general sources, and second, estimating the SORDF when only finite training data of the intrinsic state-extrinsic observation pair is available.

\section*{Acknowledgement}

The work of J.~Liu and W.~Zhang was supported in part by the National Key Research and Development Program of China under Grant 2018YFA0701603 and the Key Research Program of Frontier Sciences of CAS under Grant QYZDY-SSW-JSC003, and the work of H.~V.~Poor was supported in part by the U.S. National Science Foundation under Grant CCF-1908308.


\section*{Appendix: SORDF of General Jointly Gaussian Model}

For all $D_{\mathrm{s}} > \mathsf{mmse}$, $D_{\mathrm{a}} > 0$, the value $R(D_{\mathrm{s}}, D_{\mathrm{a}})$ of SORDF of the jointly Gaussian model has been shown in Section IV to be the solution of the optimization problem described by (13)-(15). We now proceed to solve this problem.

This problem is a Gaussian rate-distortion problem with two distortion constraints, where $[\mathbf{\Sigma}_{\rvs \rvx} \mathbf{\Sigma}_{\rvx}^{-1} \rvx, \rvx]$ is the source and $[\hat{\rvs}, \hat{\rvx}]$ is the reproduction. 

The following lemma, whose proof is an immediate consequence of the maximum entropy property of vector Gaussian distribution under a covariance constraint, is useful for our derivation.

\begin{lem}
\label{lem:minmi}
If $\rvx$ is a Gaussian random vector with zero mean and covariance matrix $\mathbf{\Sigma}_{\rvx}$, and $\hat{\rvx}$ is a random vector with the same dimension as $\rvx$, then
\begin{equation}
I(\rvx; \hat{\rvx})
\ge \frac{1}{2} \log \left(\frac{\det(\mathbf{\Sigma}_{\rvx})}{\det(\mathbf{\Delta})}\right),
\end{equation}
where $\mathbf{\Delta} = \mathbf{E}((\rvx - \hat{\rvx}) (\rvx - \hat{\rvx})^{T})$. Equality holds if and only if $\rvx - \hat{\rvx}$ and $\hat{\rvx}$ are two independent zero-mean Gaussian vectors with covariance matrices $\mathbf{\Delta}$ and $\mathbf{\Sigma}_{\rvx} - \mathbf{\Delta}$ respectively.
\end{lem}

Denote the dimension of $\rvx$ and $\hat{\rvx}$ by $m$ and the dimension of $\rvs$ and $\hat{\rvs}$ by $l$. Consider any random vectors $\hat{\rvs} \in \reals^{l \times 1}$ and $\hat{\rvx} \in \reals^{m \times 1}$ that satisfy (14) and (15). A lower bound of $I(\rvx; \hat{\rvs}, \hat{\rvx})$ then follows from Lemma \ref{lem:minmi}. Define random vector $\bar{\rvx} = \mathbf{E}(\rvx | \hat{\rvs}, \hat{\rvx})$ and matrix $\bar{\mathbf{\Delta}} = \mathbf{E}((\rvx - \bar{\rvx}) (\rvx - \bar{\rvx})^{T})$. Clearly, $\rvx \leftrightarrow (\hat{\rvs}, \hat{\rvx}) \leftrightarrow \bar{\rvx}$ form a Markov chain, so
\begin{equation}
I(\rvx; \hat{\rvs}, \hat{\rvx}) \ge I(\rvx; \bar{\rvx}).
\label{e:dpinequality} 
\end{equation}
By Lemma \ref{lem:minmi}, we have
\begin{equation}
I(\rvx; \bar{\rvx})
\ge \frac{1}{2}
\log \left(\frac{\det(\mathbf{\Sigma}_{\rvx})}{\det(\bar{\mathbf{\Delta}})}\right).
\label{e:gaussmi} 
\end{equation}

Note that $\bar{\rvx}$ is an MMSE estimate of $\rvx$, and $\mathbf{\Sigma}_{\rvs \rvx} \mathbf{\Sigma}_{\rvx}^{-1} \bar{\rvx} = \mathbf{E}(\mathbf{\Sigma}_{\rvs \rvx} \mathbf{\Sigma}_{\rvx}^{-1} \bar{\rvx} | \hat{\rvs}, \hat{\rvx})$ is an MMSE estimate of $\mathbf{\Sigma}_{\rvs \rvx} \mathbf{\Sigma}_{\rvx}^{-1} \rvx$, upon observing $(\hat{\rvs}, \hat{\rvx})$. Hence we have the following two inequalities:
\begin{align}
\mathbf{E}(\|\rvx - \bar{\rvx}\|^{2})
& \le \mathbf{E}(\|\rvx - \hat{\rvx}\|^{2}), \\
\mathbf{E}(\|
  \mathbf{\Sigma}_{\rvs \rvx} \mathbf{\Sigma}_{\rvx}^{-1} \rvx
  - \mathbf{\Sigma}_{\rvs \rvx} \mathbf{\Sigma}_{\rvx}^{-1} \bar{\rvx}
\|^{2})
& \le \mathbf{E}(\|\mathbf{\Sigma}_{\rvs \rvx} \mathbf{\Sigma}_{\rvx}^{-1} \rvx - \hat{\rvs}\|^{2}).
\end{align}
On the other hand, $\mathbf{E}(\|\rvx - \bar{\rvx}\|^{2}) = \tr(\bar{\mathbf{\Delta}})$ and $\mathbf{E}(\|\mathbf{\Sigma}_{\rvs \rvx} \mathbf{\Sigma}_{\rvx}^{-1} \rvx - \mathbf{\Sigma}_{\rvs \rvx} \mathbf{\Sigma}_{\rvx}^{-1} \bar{\rvx}\|^{2}) = \tr(\mathbf{\Sigma}_{\rvs \rvx} \mathbf{\Sigma}_{\rvx}^{-1} \bar{\mathbf{\Delta}} \mathbf{\Sigma}_{\rvx}^{-1} \mathbf{\Sigma}_{\rvs \rvx})$. Along with (14) and (15), we thus get
\begin{align}
\tr(\bar{\mathbf{\Delta}}) & \le D_{\mathrm{a}},\\
\tr(\mathbf{\Sigma}_{\rvs \rvx} \mathbf{\Sigma}_{\rvx}^{-1} \bar{\mathbf{\Delta}} \mathbf{\Sigma}_{\rvx}^{-1} \mathbf{\Sigma}_{\rvs \rvx}^{T}) & \le D_{\mathrm{s}} - \mathsf{mmse}.
\end{align}

Note that both $\bar{\mathbf{\Delta}}$ and $\mathbf{\Sigma}_{\rvx} - \bar{\mathbf{\Delta}}$ are positive semidefinite, and we can define the set
\begin{align}
A(D_{\mathrm{s}}, D_{\mathrm{a}})
= \{
  \mathbf{\Delta} \in \reals^{m \times m}
|
  \mathbf{0}_{m \times m} \le \mathbf{\Delta} \le \mathbf{\Sigma}_{\rvx}, \nonumber\\
  \tr(\mathbf{\Sigma}_{\rvs \rvx} \mathbf{\Sigma}_{\rvx}^{-1} \mathbf{\Delta} \mathbf{\Sigma}_{\rvx}^{-1} \mathbf{\Sigma}_{\rvs \rvx}^{T})
  \le D_{\mathrm{s}} - \mathsf{mmse},
  \tr(\mathbf{\Delta}) \le D_{\mathrm{a}}
\},
\end{align}
which $\bar{\mathbf{\Delta}}$ should belong to.

From \eqref{e:dpinequality} and \eqref{e:gaussmi}, we obtain the lower bound
\begin{equation}
I(\rvx; \hat{\rvs}, \hat{\rvx})
\ge \min_{\mathbf{\Delta} \in A(D_{\mathrm{s}}, D_{\mathrm{a}})} \frac{1}{2}
  \log \left(\frac{\det(\mathbf{\Sigma}_{\rvx})}{\det(\mathbf{\Delta})}\right).
\label{e:lbound} 
\end{equation}

The lower bound \eqref{e:lbound} is in fact achievable. To see this, let $\mathbf{\Delta}^{*}$ be the solution of the optimization of \eqref{e:lbound}, and $\hat{\rvx}$ and $\rvz$ be independent zero-mean Gaussian vectors with covariance matrices $\mathbf{\Sigma}_{\rvx} - \mathbf{\Delta}^{*}$ and $\mathbf{\Delta}^{*}$ respectively. Define $\rvx = \hat{\rvx} + \rvz$ and $\hat{\rvs} = \mathbf{\Sigma}_{\rvs \rvx} \mathbf{\Sigma}_{\rvx}^{-1} \hat{\rvx}$. So $\rvx \sim \mathcal{N}(\mathbf{0}_{m \times 1}, \mathbf{\Sigma}_{\rvx})$ and $I(\rvx; \hat{\rvs}, \hat{\rvx}) = I(\rvx; \hat{\rvx})$. Straightforward calculations yield $\mathbf{E}(\|\mathbf{\Sigma}_{\rvs \rvx} \mathbf{\Sigma}_{\rvx}^{-1} \rvx - \hat{\rvs}\|^{2}) = \tr(\mathbf{\Sigma}_{\rvs \rvx} \mathbf{\Sigma}_{\rvx}^{-1} \mathbf{\Delta}^{*} \mathbf{\Sigma}_{\rvx}^{-1} \mathbf{\Sigma}_{\rvs \rvx}^{T})$ and $\mathbf{E}(\|\rvx - \hat{\rvx}\|^{2}) = \tr(\mathbf{\Delta}^{*})$, so the distortion constraints (14) and (15) are satisfied. By Lemma \ref{lem:minmi}, we have that equality in \eqref{e:lbound} is achieved by $\mathbf{\Delta}^{*}$.

In conclusion, the SORDF of the jointly Gaussian model is given by
\begin{equation}
R(D_{\mathrm{s}}, D_{\mathrm{a}})
= \min_{\mathbf{\Delta} \in A(D_{\mathrm{s}}, D_{\mathrm{a}})} \frac{1}{2}
\log \left(\frac{\det(\mathbf{\Sigma}_{\rvx})}{\det(\mathbf{\Delta})}\right).
\end{equation}
A variety of software libraries are available to solve the semidefinite programming of $R(D_{\mathrm{s}}, D_{\mathrm{a}})$.

\begin{figure}[ht]
\centering
\includegraphics[width = 0.5\textwidth]{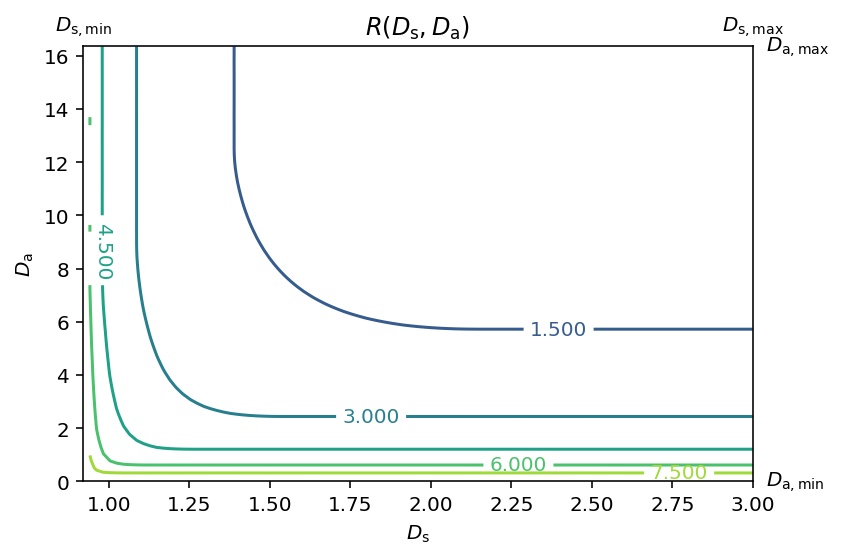}
\caption{Contour plot of SORDF in Example \ref{xmp:gvector}. $D_{\mathrm{s, min}} = \mathsf{mmse}$, $D_{\mathrm{s, max}} = \tr(\mathbf{\Sigma}_{\rvs})$, $D_{\mathrm{a, min}} = 0$, and $D_{\mathrm{a, max}} = \tr(\mathbf{H} \mathbf{\Sigma}_{\rvs} \mathbf{H}^{T} + \mathbf{\Sigma}_{\rvw})$.}
\label{f:dads} 
\end{figure}

It can be verified that the solution of the case (19) in Section IV-C is consistent with our general solution here. Below is another example which cannot be manually solved in closed form.
\begin{example}
\label{xmp:gvector} 
Let $\rvs$ and $\rvw$ be independent zero-mean Gaussian vectors with covariance matrices $\mathbf{\Sigma}_{\rvs}$ and $\mathbf{\Sigma}_{\rvw}$ respectively, and define $\rvx = \mathbf{H} \rvs + \rvw$, where
\begin{equation*}
\mathbf{\Sigma}_{\rvs} = \begin{pmatrix}
  1 & \\
  & 2
\end{pmatrix},
\mathbf{\Sigma}_{\rvw} = \begin{pmatrix}
  10 & & \\
  & 1 & \\
  & & 0.1
\end{pmatrix},
\mathbf{H} = \begin{pmatrix}
  1 & 0 \\
  0 & - 1 \\
  0.5 & 1
\end{pmatrix}.
\end{equation*}
So $(\rvs, \rvx)$ is the Gaussian semantic source, and we compute its SORDF by CVXPY, as shown in Figure \ref{f:dads}.
\end{example}

\end{document}